\documentclass{elsart}
\usepackage{epsfig}

\newcommand {\be}{\begin{equation}}
\newcommand {\ee}{\end{equation}}
\newcommand{\bea}{\begin{eqnarray}}
\newcommand{\eea}{\end{eqnarray}}
\newcommand {\vett}[1] {\mathbf{#1}}

\catcode `\@=11
\@addtoreset{equation}{section}
\def\theequation{\arabic{section}.\arabic{equation}}
\catcode `\@=12     
\begin{document}
\begin{frontmatter}

\title{Schematic phase diagram and collective excitations in the
collisional regime for trapped boson-fermion mixtures at zero temperature}
\author[Pisa]{A. Minguzzi\thanksref{CA}} and
\author[Pisa]{M. P. Tosi} 
\address[Pisa]{Istituto Nazionale di Fisica della Materia and Classe
di Scienze,  Scuola Normale Superiore, Piazza dei Cavalieri 7, I-56126
Pisa, Italy} 
\thanks[CA]{Corresponding author. Tel.: +39 50 509058; fax: +39 50
563513; e-mail: minguzzi@cibs.sns.it.} 

\begin{abstract}
We discuss the ground state and the small-amplitude
excitations of dilute boson-fermion mixtures confined in spherical
harmonic traps at $T = 0$, assuming repulsive boson-boson interactions
and with each component being in a single hyperfine state. From
previous studies of the equilibrium density profiles we propose a
schematic phase diagram in a plane defined by the variables
$a_{bf}/a_{bb}$ and $a_{bb}k_f^{(0)}$, 
where $a_{bb}$ and $a_{bf}$ are the boson-boson and boson-fermion scattering
lengths and $k_f^{(0)}$ is the Fermi wave number at the centre of the
trap. With this background we turn to the equations of motion for
density fluctuations in the collisional regime and discuss some
general features of the eigenmodes. We display analytic solutions for
sound waves in a quasi-homogeneous mixture and for surface modes at
weak fermion-boson coupling. 

\end{abstract}

\end{frontmatter}

\section{Introduction}

The achievement of Bose-Einstein condensation in trapped atomic gases
[1 - 3] has given new impulse to the study of dilute quantal
fluids. Experimental attention has extended to the properties of
double Bose condensates [4, 5] and more recently to the attainment of
quantal degeneracy in gases of fermionic atoms [6 - 10] and in dilute
mixtures of bosons and fermions [11]. A specific motivation for
interest in boson-fermion mixtures is that the collisions between
fermions and bosons can foster thermalization of the fermionic
component and induce so-called sympathetic cooling [4, 12]. Elastic
collisions between fermions in a single hyperfine state are inhibited
by the Pauli principle [13] and even in multi-component Fermi gases
the collisions between atoms in different hyperfine states become
ineffective at very low temperature because of Fermi factors [9]. 

A number of theoretical studies have already been addressed to
properties of dilute boson-fermion mixtures confined in harmonic
traps. They have concerned the equilibrium density profiles of the two
species both at zero and at finite temperature [14 - 17], the kinetic
energy of the fermionic component [18] and collective excitations from
sum rules in the collisionless regime [19]. Our main focus here is on
small-amplitude collective excitations for a mixture under harmonic
confinement in the collisional regime. For a detailed discussion on
how this regime may be achieved in dilute Fermi gases we refer to the
work of Amoruso {\it et al.}~[20]. 
	
We first set up a general background for our discussion of collective
excitations in boson-fermion mixtures by constructing a schematic
phase diagram from the earlier studies of the equilibrium state [14 -
18]. This helps to identify various dynamical behaviours, which may
later be examined in detail by explicit numerical solution of the
equations of motion. We then report analytic results for the
eigenfrequencies of the mixture in limiting situations, corresponding
to mixing of Bogolubov sound and fermionic first sound in a nearly
homogeneous mixture and to surface modes in a regime of weak
fermion-boson coupling. 

We assume throughout a positive boson-boson scattering length, as is
applicable to Bose condensates of $^{87}$Rb and $^{23}$Na. Mixtures of
$^6$Li  and
$^7$Li should have special interest, once condensation is realized for
$^7$Li in a hyperfine state where the scattering length is
positive. According to calculations on the collisional properties of
ultracold potassium [21], the $^{40}$K-$^{41}$K mixture should also be an
interesting system for study from the present viewpoint, if
condensation of the rare isotope $^{41}$K can be achieved. 

\section{Equilibrium density profiles and phase diagram}

We consider a system of $N_f$ fermions of mass $m_f$ and $N_b$ bosons of mass
$m_b$ at $T = 0$, confined by spherically symmetric external potentials
$V_{f,b}(r)=m_{f,b} \omega_{f,b}^2r^2/2$
with frequencies $\omega_f$ and $\omega_b$. We assume that a single
spin state is 
trapped for each component of the mixture and that all bosons are in
the condensate. We can then omit the fermion-fermion interaction,
which is inhibited by the Pauli exclusion principle. The boson-boson
and boson-fermion interactions are described by contact potentials
with scattering lengths $a_{bb}>0$ and $a_{bf}$, yielding coupling-strength
parameters  $g=4 \pi \hbar^2 a_{bb}/m_b > 0$ and $f=2 \pi
\hbar^2 a_{bf}/m_r$ where $m_r$ is the reduced boson-fermion mass. 
	
The equations of motion for the partial particle densities
$n_\sigma(\vett r,t)$ and
partial current densities $\vett j_\sigma(\vett r,t)$, where $\sigma$
is an index for the atomic 
species ($\sigma = f$ or $b$) follow from those for the one-body density
matrix upon projection on the main diagonal (see for instance
[22]). They take the form of continuity equations, 
\be
\partial_t n_\sigma (\vett r,t)=-\nabla \cdot \vett j_\sigma (\vett r,t)
\ee
and of generalized hydrodynamic equations which, in a mean-field
approximation and neglecting coupling to energy fluctuations, can be
written as 
\be
m_f \partial_t \vett j_f(\vett r,t)=-\nabla \cdot \Pi^{(f)}(\vett
r,t)-n_f(\vett r,t) \nabla\left[V_f(\vett r)+f n_b(\vett r,t)-\mu_f\right]
\ee
and
\bea
m_b \partial_t \vett j_b(\vett r,t)&=&-\nabla \cdot \Pi^{(b)}(\vett
r,t)\nonumber \\ &&-n_b(\vett r,t) \nabla\left[V_b(\vett r)+g n_b(\vett r,t)+f
n_f(\vett r,t)-\mu_b\right] 
\eea
In these equations $\Pi^{(\sigma)}(\vett r,t)$ are the kinetic stress
tensors and $\mu_\sigma$ are the 
chemical potentials of the two atomic species. In the following we
treat the kinetic stress tensors by a local density approximation for
fermions,
\be
\Pi_{ij}^{(f)}(\vett r,t)=\frac 2 5 A [n_f(\vett r,t)]^{5/3} \delta_{ij}
\ee
with $A=\hbar^2 (6 \pi^2)^{2/3}/2 m_f$ and by a Thomas-Fermi
approximation for bosons, 
\be
\Pi^{(b)}(\vett r,t)=0\;.
\ee
In these equations we have dropped terms which are quadratic in the
velocity field.

The treatment given above assumes that local equilibrium has been
established for the ground-state density profiles and that it is
maintained during dynamic fluctuations of the particle densities. The
dynamical behaviour that we shall discuss in Sect.3 will therefore
correspond to a collisional regime [20].

\subsection{ Equilibrium density profiles}

The equilibrium density profiles ($n_\sigma^{eq}(r)$, say) follow from
Eqs. (2.2) - 
(2.5) by setting the current densities to zero. The results are well
known [14, 15]:
\be
n_b^{eq}(r)=g^{-1}\left[\mu_b-V_b(r)-f
n_f^{eq}(r)\right]\theta(\mu_b-V_b(r)-f n_f^{eq}(r)) 
\ee
and
\be
n_f^{eq}(r)=A^{-3/2}\left[\mu_f-V_f(r)-f
n_b^{eq}(r)\right]^{3/2}\theta(\mu_f-V_f(r)-f n_b^{eq}(r))  \;.
\ee
The chemical potentials are to be determined by self-consistency
conditions on the particles numbers, $N_\sigma=\int
n_\sigma^{eq}(r)d\vett r$. On comparing the two density
profiles in the limit of vanishing boson-fermion coupling, it is
easily seen that the quantity $(2A/3)[n_f^{eq}(r)]^{-1/3}$ may be
viewed as an effective 
fermion-fermion coupling arising from the kinetic pressure of the
Fermi gas [23]. It is also useful to introduce the radii $R_b=(2
\mu_b/m_b \omega_b^2)^{1/2}$ and $R_f=(2 \mu_f/m_f\omega_f^2)^{1/2}$ for
the two clouds in the absence of interactions, but taken at the true
values of the chemical potentials.

For values of the boson-boson coupling and of the confinement
frequencies which are typical of current experiments, and assuming
comparable magnitudes for the boson-boson and the boson-fermion
coupling, the fermionic cloud is considerably more dilute than the
bosonic cloud. In this situation the boson cloud is essentially
unaffected by the interactions with the fermionic component and the
latter can be treated as confined inside a double-parabola effective
potential $V^{eff}_f(r)$ [15]. This is given by
\be
V_f^{eff}(r)=\left\{\begin{array}{l r}
m_f \omega_f^2(1-\gamma) r^2/2+f \mu_b/g & {\rm for} \; r<R_b \\
m_f \omega_f^2 r^2/2 & {\rm for} \; r>R_b
\end{array}\right.
\ee
where
\be
\gamma=(f m_b \omega_b^2)/(g m_f \omega_f^2)\;.
\ee	
Evidently, if $\gamma>1$ the effective potential (2.8) has negative concavity
inside the Bose radius and the fermions will tend to be expelled from
the centre of the trap. The result at strong coupling is phase
separation in the mixture, with the fermions forming a shell around
the boson cloud [15].

A second type of phase separation for the boson-fermion mixture was
demonstrated in the work of M{\o }lmer [14]. In this case one takes
$N_f \simeq N_b$ and
considers very large values of the boson-boson coupling $g$, such that
the two clouds acquire similar densities. One ultimately finds that
the bosons are expelled from the centre of the trap and form a shell
around the fermion cloud.

Finally, Eq. (2.8) is valid also in the case $\gamma<0$, {\it i.e.}
for attractive 
boson-fermion interactions, as long as the fermionic cloud is more
dilute than the boson cloud. However, the fermions now tend to draw
the bosons towards the centre of the trap, ultimately leading to
collapse once $\gamma$ becomes sufficiently negative [24]. 

\subsection{ Schematic phase diagram at zero temperature}

As shown by Vichi {\it et al.} [18], the thermodynamics of the boson-fermion
mixture at $T = 0$ is well characterized by means of two scaling
parameters, which are $\gamma$ in Eq. (2.9) and
\be
x=\left(\frac{R_b}{R_f}\right)^{1/2}=\left(\frac{m_f \omega_f}{2 m_b \omega_b}\right)^{1/2} \left[15\frac{a_{bb}}{a_{ho}}\frac{N_b}{(6N_f)^{5/6}}\right]^{1/5}\;,
\ee
with $a_{ho}=(\hbar/m_b\omega_b)^{1/2}$. The description of the system
with only two scaling 
parameters, instead of the eight original ones entering the expression
of the equilibrium density profiles, is a major simplification. The
ratio $R_b/R_f$ determines the deviation of the kinetic energy of the fermion
cloud from its ideal-gas value [18].

For the purposes of the present discussion we make the simplifying
assumptions $m_f=m_b$, $\omega_f=\omega_b$ and $N_f=N_b$. In this case
the two scaling parameters 
take the simple expressions $x=(15 a_{bb}k_f^{(0)}/48)^{1/5}$ and
$\gamma=a_{bf}/a_{bb}$, with $k_f^{(0)}=(48 N_f)^{1/6}/a_{ho}$ being
the Fermi wave 
number at the trap centre for a non-interacting Fermi gas. As remarked
under Eq. (2.7) above, the quantity $a_{ff}\equiv 3 \pi/4 k_f^{(0)}$
may be viewed as an effective 
fermion-fermion scattering length arising from the Pauli kinetic
pressure. We can now proceed to construct a schematic phase diagram
for the boson-fermion mixture in the plane defined by the variables
$\gamma=a_{bf}/a_{bb}$ 
and $y=a_{bb}k_f^{(0)}$. Figure 1 shows the result for the case
$\gamma>0$ ({\it i.e.}~repulsive
boson-fermion interactions).

\begin{figure}
\centerline{\epsfig{figure=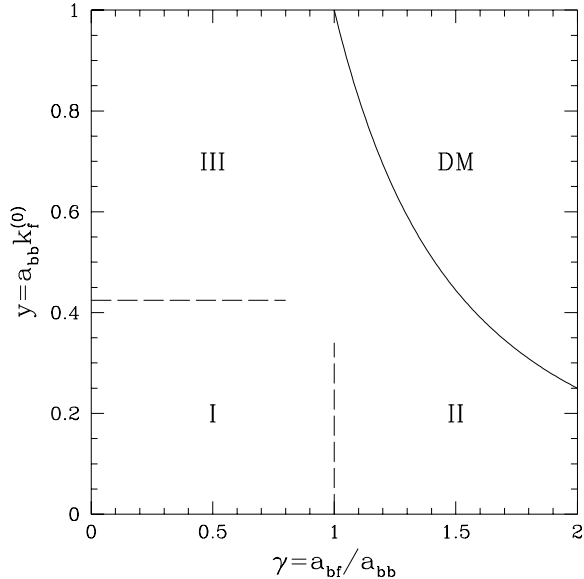,width=8cm}}
\caption{Schematic phase diagram of boson-fermion mixtures in a spherical
harmonic trap, for the case of repulsive boson-boson and boson-fermion
interactions. The phase-separation region is marked DM (for
"demixed"). See the main text for a description of areas I, II and III
in the miscibility region.}
\end{figure}

We search for the boundary of phase separation by means of a simple
condition of linear stability on the two-by-two matrix of scattering
lengths $a_{bb}$, $a_{bf}$ and $a_{ff}$ [23] (see also [25]). A mixed
state is stable if 
the inequality $a_{bb}a_{ff}-a_{bf}^2>0$ holds, {\it i.e.}~when
\be
k_f^{(0)}a_{bb}\le \frac{3\pi}{4}\left(\frac{a_{bb}}{a_{bf}}\right)^2
\ee
Within this region we can distinguish a region of appreciable overlap
between the two clouds (region I in Figure 1), bounded by two regions
of diminishing overlap: a region at $\gamma>1$ where the fermions are being
expelled from the centre of the trap (region II in Figure 1) and by a
region at $y>4/3\pi$ where the bosons are being expelled from the centre of
the trap (region III in Figure 1). If the spherical symmetry of the
trapping potential is preserved in the demixed state, these two
regions end into spherical-shell structures with the fermions at large
$\gamma$ (or the bosons at large $y$) forming the outer shell. 

\section{ Small amplitude oscillations around the equilibrium profiles}

The equations of motion for small-amplitude density fluctuations
$\tilde n_f(\vett r,t)$
and $\tilde n_b(\vett r,t)$ in the collisional regime are found by
combining Eq. (2.1) with 
the linearized form of Eqs. (2.2) and (2.3). They are
\be
\partial_t^2\tilde n_f(\vett r,t)=\frac{1}{m_f} \nabla \cdot
\left\{n_f^{eq}(r)\nabla\left[\frac  2 3 A(n_f^{eq}(
r))^{-1/3} \tilde n_f(\vett r,t)+f\tilde n_b(\vett r,t) \right]\right\}
\ee
and
\be
\partial_t^2  \tilde n_b(\vett r,t)= \frac{1}{m_b} \nabla \cdot
\left\{n_b^{eq}(r)\nabla\left[ g\tilde n_b(\vett r,t)+f \tilde
n_f(\vett r,t)\right]\right\}\;. 
\ee
Again the quantity $(2A/3)[n_f^{eq}(r)]^{-1/3}$ enters Eq. (3.1) as an
effective fermion-fermion 
coupling.  

It is easily checked that the form of Eqs. (3.1) and (3.2) is such as
to satisfy the generalized Kohn theorem [26, 27] for the
centre-of-mass coordinate $\vett x(t)$ of the whole fluid when the two
confinement frequencies coincide. In this mode of motion we have
$n_{b,f}(\vett r,t)=n_{b,f}^{eq}(\vett r-\vett x(t))$
and hence
\be
\tilde n_{b,f}(\vett r,t)=-\vett x (t)\cdot \nabla n_{b,f}^{eq}(r)
\ee
for small amplitude oscillations. By substituting Eq. (3.3) in
Eqs. (3.1) and (3.2) and using Eqs. (2.6) and (2.7), we find
\be
m_{b,f}\partial^2_t \vett x(t)\cdot \nabla n_{b,f}^{eq}(
r)=\sum_{i,j}x_j(t)\nabla_i\left[ n_{b,f}^{eq}(
r)\nabla_i\nabla_jV_{b,f}(r)\right]\;. 
\ee
Therefore, the centre of mass of the system oscillates at the bare
trap frequency.

Evidently, a general solution of Eqs. (3.1) and (3.2) can only be
obtained numerically, since the equilibrium density profiles are not
known analytically from Eqs. (2.6) and (2.7). The dynamics of the
fluid as a function of the scaling parameters will be especially rich
in the regions of the phase diagram in which the nature of the partial
density profiles is changing on the approach to phase separation. Of
course, this will be signalled by a softening of the modes associated
with concentration fluctuations. Below we restrict ourselves to
analytic results which are applicable to special regimes.

\subsection{ Homogeneous limit}

The dynamics of a spatially homogeneous mixture is relevant to the
situation in which the density profiles of both components are slowly
varying in space, on the length scale set by the mean interparticle
distances. Taking the equilibrium densities as constant, it is easily
seen from Eqs. (3.1) and (3.2) that the eigenmodes are two sound
waves, with linear dispersion relation $\omega_{1,2}(q)=c_{1,2}
q$. The sound velocities are 
given by
\be
c_{1,2}^2=\frac 1 2 \left[c_b^2+c_f^2\pm
\sqrt{(c_b^2-c_f^2)^2+4f^2n_b^{eq}n_f^{eq}/m_bm_f}\right]\;. 
\ee
In Eq. (3.5) $c_b=(g n_b^{eq}/m_b)^{1/2}$ is the speed of Bogolubov
sound and $c_f=[2 A(n_f^{eq})^{2/3}/3m_f]^{1/2}$ is that of
first sound in a Fermi gas.

This result may be useful in regard to measurements of the speed of
sound in elongated systems, in experiments such as already carried out
by Matthews {\it et al.}~[30] on a single Bose condensate.

\subsection{ Surface modes for weak boson-fermion coupling}

Assuming equal confinement frequencies for the two components of the
mixture ($\omega_{ho}$, say), it is known from earlier results of
Stringari [28] 
for a Bose condensate and of Amoruso {\it et al.}~[29] for a one-component
Fermi gas that in the limit $f = 0$ the frequencies of the surface modes
($n $= 0) with angular momentum number  are equal and given by
$\omega_l=l^{1/2}\omega_{ho}$.  A
perturbative treatment of the effect of fermion-boson coupling is
therefore especially simple for these modes.

We define the unperturbed equilibrium density profiles
$n_{b,f}^{(0)}(r)$ and the 
$\vett r$-dependent amplitudes $\tilde n_{b,f}^{(l,m)}(\vett r)$ of
the unperturbed density fluctuations, for 
each value of the angular momentum $l$ and of its $z$-component $m$. After
Fourier-trans\-for\-ming Eqs. (3.1) and (3.2) and linearizing them in the
coupling parameter $f$, we multiply Eq. (3.1) by
$[n_f^{(0)}(r)]^{-1/3}\tilde n_f^{(l,m)}(\vett r)$ and Eq. (3.2) by $\tilde
n_b^{(l,m)}(\vett r)$, 
and then integrate both equations over $\vett r$. We obtain in this way a
determinantal equation for the shift of the eigenmode frequencies $\Omega_l$
due to the fermion-boson coupling. Writing
\be
\Omega_l^2=\omega_{ho}^2(l+\Delta_l)\;,
\ee
we find
\be
\Delta_l=\frac{a_{bf}}{2a_{ho}}\left[-(E+B)\pm {\rm
sign}(a_{bf})\sqrt{(E-B)^2+4CD}\right] \;,
\ee
this result being valid for either sign of $a_{bf}$. The quantities entering
Eq. (6) are defined in the Appendix, where it is also shown that they
can all be expressed in terms of the Gauss hypergeometric function
[31].

The result given in Eq. (3.7) should be better applicable in the case
$R_b<R_f$, since in deriving it we have assumed that the equilibrium density
profiles for both components are perturbed in a symmetric manner. This
corresponds to working in region I of the phase diagram (and in the
corresponding region at negative $\gamma$).

\section{ Summary and discussion}

The main features of the zero-temperature phase diagram that we have
sket\-ched in Figure 1 for boson-fermion mixtures arise from the
competition between the kinetic energy of the Fermi gas and the
repulsive boson-boson and boson-fermion interactions. The former
disfavours localization and phase separation, while the latter favour
the spatial separation of the two components. On varying these system
parameters while preserving the spherical symmetry of the trapped
fluid mixture, a spontaneous symmetry breaking occurs towards a
situation where the two components are spatially separated along the
radial direction. Evidently, axial separation of the two components
would instead be observed in an anisotropic trap. One may think of
exploiting Feshbach resonances to tune the scattering lengths towards
such phase-separation regime.

Given the complex behaviour of the density profiles on the approach to
phase separation, a complete investigation of the oscillatory
eigenmodes of a boson-fermion mixture can only be carried out by
numerical means. This will be well worth doing once a mixture of
specific experimental relevance and with reasonably known scattering
lengths is identified. Of course, the dynamics in the phase-separated
region is simply related to that of the two pure components, aside for
the presence of interfacial modes. On the other hand, we have seen
that suggestive analytic results can be obtained in the
quasi-homogeneous limit and in the weak boson-coupling regime, for
either sign of the boson-fermion coupling. With increasingly large
attractions between boson and fer\-mions, however, the mixture is
expected to undergo collapse [24].

\ack

We thank Dr Ilaria Meccoli for many useful discussions and for her
help in the early stages of this work. 

\catcode `\@=11
\@addtoreset{equation}{section}
\def\theequation{\Alph{section}.\arabic{equation}}
\catcode `\@=12  

\appendix

\section{Calculation of frequency shifts for surface modes}
We introduce the notations $|F\rangle=\tilde n_f^{(l,m)}(\vett r)$ and
$|B\rangle=\tilde n_b^{(l,m)}(\vett r)$, define the operators
$P_{b,f}=(f/m_{b,f})\nabla \cdot [n_{b,f}^{(0)}(r)\nabla]$ and
$D_f=(-f/m_f) \nabla \cdot \{(n_f^{(0)}(r))^{1/3} \nabla
[(n_f^{(0)}(r))^{-1/3}]\}$, 
and introduce the scalar-product notation $\langle F|\hat O
|F\rangle=\int \tilde n_f^{(l,m)}(\vett r)\hat O n_f^{(l,m)}(\vett
r)d\vett r$ etcetera. With these
notations we find the following expressions for the quantities
entering Eq. (3.7): 
\bea
B&=&\langle
F|(n_f^{(0)}(r))^{-1/3}D_f|F\rangle/\langle F|(n_f^{(0)}(r))^{-1/3}|F
\rangle \nonumber \\ &=&\frac{2
\pi^{-1/2}l}{\alpha_{bb}X_f^{2l+2}}\frac{\Gamma(l+3)}{\Gamma(l+3/2)}
\int_0^{X_b}dx\,x^{2l+2} (X_f^2-x^2)^{-1/2}(2 X_f^2+X_b^2-3x^2),
\eea
\bea
C&=&\langle
F|(n_f^{(0)}(r))^{-1/3}P_f|B\rangle/\langle F|(n_f^{(0)}(r))^{-1/3}|F
\rangle \nonumber \\
&=&\frac{-2l}{\pi^{5/2}X_f^{2l+2}}\frac{\Gamma(l+3)}{\Gamma(l+3/2)} 
\int_0^{X_b}dx\,x^{2l+2} (X_f^2-x^2)^{1/2}\;,
\eea
\bea
D&=&\langle
B|P_b|F\rangle/\langle B|B
\rangle \nonumber \\ &=&\frac{-(2l+3)}{2\alpha_{bb}X_b^{2l+3}}
\int_0^{X_b}dx\,x^{2l+2}\left\{2l (X_f^2-x^2)^{1/2}+X_f^2
(X_f^2-x^2)^{-3/2}(X_b^2-x^2)\right.\nonumber \\&&\left.+(X_f^2-x^2)^{-1/2}[-x^2+2(l+1)(X_b^2-x^2)]
\right\}\;,
\eea
and
\bea
E&=&-\langle B|P_f|B\rangle/\langle B|B
\rangle \nonumber \\ &=&\frac{l(2l+3)}{2\pi^2X_b^{2l+3}}
\int_0^{X_b}dx\,x^{2l+2}(X_f^2-x^2)^{1/2}\;.
\eea
In these equations $\Gamma(n)$ is the Gamma function and we have set
$X_{b,f}=R_{b,f}/a_{ho}$ and $\alpha_{bb}=a_{bb}/a_{ho}$. 
	
All the integrals in Eqs. (A.1) -(A.4) can be expressed through the
Gauss hypergeometric function $_2F_1(a,b;c;z)$, using the relation
[31]
\be
_2F_1\left(\frac{1+\alpha}{2},-\frac{n}{2};\frac{3+\alpha}{2};
\frac{X_b^2}{X_f^2}\right)
=\frac{1+\alpha}{X_f^nX_b^{\alpha+1}}\int_0^{X_b}  
dx \,x^\alpha (X_f^2-x^2)^{n/2}\;.
\ee

\end{document}